\documentclass[prd,aps,showpacs,epsf,floats]{revtex4}

\usepackage{amssymb}
\usepackage{amsfonts}
\usepackage{amsmath}
\usepackage{graphicx}

\begin{document}

\title{Quantum Stabilizer Codes Embedding Qubits Into Qudits}
\author{Carlo Cafaro$^{1\text{, }2\text{, }3}$, Federico Maiolini$^{1}$ and
Stefano Mancini$^{1}$}
\affiliation{$^{1}$School of Science and Technology, Physics Division, University of
Camerino, I-62032 Camerino, Italy}
\affiliation{$^{2}$Optical Quantum Information Theory Group, Max-Planck Institute for the
Science of Light, Gunther-Scharowsky-Str.1/Bau 26 and }
\affiliation{$^{3}$Institute of Theoretical Physics, Universitat Erlangen-Nurnberg,
Staudtstr. 7/B2, 91058 Erlangen, Germany}

\begin{abstract}
We study, by means of the stabilizer formalism, a quantum error correcting
code which is alternative to the standard block codes since it embeds a
qubit into a qudit. The code exploits the non-commutative geometry of
discrete phase space to protect the qubit against both amplitude and phase
errors. The performance of such code is evaluated on Weyl channels by means
of the entanglement fidelity as function of the error probability. A
comparison with standard block codes, like five and seven qubit stabilizer
codes, shows its superiority.
\end{abstract}

\maketitle

\section{Introduction}

In order to describe realistic quantum information processes, quantum errors
induced by environmental noise must be taken into account. This can be
accomplished by introducing the notion of quantum channels, that is to say
maps on the set of states of the system that are completely positive and
trace preserving \cite{petz}. At the same time, one would combat quantum
errors to avoid their detrimental effect on quantum information processes.
To this end, the method of error correcting codes has been borrowed from
classical information theory (for a comprehensive introduction to the
quantum theory of error correcting codes we refer to \cite{gotty1}). The
underlying idea is to exploit redundancy, that is to encode information in
linear subspaces (codes) of the total complex Hilbert space in such a way
that errors induced by the interaction with the environment can be detected
and corrected.

Usually a logical qubit (a two dimensional complex Hilbert space $\mathcal{H}%
_{2}$) is encoded into $n$ physical qubits (a $2^{n}$ dimensional complex
Hilbert space $\mathcal{H}_{2}^{\otimes n}$). This kind of encoding is known
as \emph{block-encoding } 
\begin{equation*}
\mathcal{H}_{2}\ni\left\vert q_{\text{logical}}\right\rangle \longmapsto
\left\vert q_{\text{physical}}\right\rangle \in\mathcal{H}_{2}^{\otimes n}%
\text{.}  \label{1}
\end{equation*}
However, there is also the possibility of \emph{embedding} a logical qubit
into a $d$-dimensional quantum physical system, i.e. a \emph{qudit} with
complex Hilbert space $\mathcal{H}_{d}\neq\mathcal{H}_{2}^{\otimes n}$. We
refer to this kind of encoding as \emph{embedding} or \emph{qudit-encoding} 
\begin{equation*}
\mathcal{H}_{2}\ni\left\vert q_{\text{logical}}\right\rangle \longmapsto
\left\vert q_{\text{physical}}\right\rangle \in\mathcal{H}_{d}\neq \mathcal{H%
}_{2}^{\otimes n}\text{,}  \label{2}
\end{equation*}
where $d\neq2^{n}$.

For block-coding schemes a powerful formalism, named stabilizer formalism 
\cite{daniel-thesis}, has been developed describing one of the most
important classes of quantum codes, namely the quantum version of linear
codes in classical coding theory. The stabilizer formalism can be extended
over non binary codes \cite{ketkar} and can be also useful for describing
embedding (qudit) codes. Actually, the idea of embedding a qubit into a
larger space without resorting to block codes was put forward in \cite%
{gotty-preskill} where a qubit was encoded into a bosonic mode (infinite
dimensional complex Hilbert space) just using the stabilizer formalism.
Later on the possibility of qudit encoding was pointed out in Ref. \cite%
{pirandola} by using the same formalism. There, being in $\mathcal{H}_{d}$,
the errors were considered as a generalization of Pauli operators,
representing shift errors ($X$-type) or phase errors ($Z$-type) or their
combination ($XZ$-type). However, in \cite{pirandola} the proposed code was
essentially classical since only $Z$-type errors were taken into
consideration.

In this article we upgrade such a coding scheme to be fully quantum, thus
able to correct $X$-type, $Z$-type and, $XZ$-type errors. We then test its
effectiveness for $d=18$ and $d=50$ on a Weyl quantum noisy channel \cite%
{amosov} (an error model characterized by errors of $X$, $Z$ and $XZ$%
-types). We allow for the possibility of considering $X$ and $Z$ errors
occurring with both symmetric and asymmetric probabilities. Finally, we
compare the performance of such qudit coding schemes to those of the
conventional five-qubit stabilizer code $[[5$, $1$, $3]]$ \cite{laflamme,
bennett} and seven-qubit stabilizer CSS (Calderbank-Shor-Steane) code $\left[
\left[ 7\text{, }1\text{, }3\right] \right] $ \cite{steane, calderbank}. We
characterize the performances of these codes by means of the entanglement
fidelity \cite{schumy}, rather than the averaged input-output fidelity used
in \cite{pirandola}. We show that the use of qudit codes may allow to save
space resources while achieving the same performance of block codes.

The layout of the article is as follows. In Section II, we briefly describe
block-encoding and then introduce qudit encoding. Special focus is devoted
to both Pauli groups of $n$-qubit vectors and generalized Pauli groups of
error operators acting on a qudit. In Section III, the Weyl noisy quantum
channel is discussed together with the entanglement fidelity. In Section IV,
we study the performance of qudit codes. In Section V, for the sake of
comparison, we quantify the performance of relevant block codes. Our final
remarks appear in Section VI.

%%%%%%%%%%%%%%%%%%%%%%%%%%%%%%%%%%%%%%%%%%%%%%%%%%%%%%%%%

\section{From block codes to embedding codes}

In this Section, we briefly recall the block-encoding error correction
schemes in terms of the stabilizer formalism. Then, using the same formalism
we introduce the qudit-encoding (embedding qubits into qudits) scheme. In
both cases we restrict our attention to a single encoded qubit.

\subsection{Pauli group of n-qubits and block-encoding}

A qubit is a two-dimensional quantum system with associated complex Hilbert
space $\mathcal{H}_{2}$. Let $\{|0\rangle,|1\rangle\}$ be the canonical
basis of this space and consider on it the Pauli operators $X\overset{\text{%
def}}{=}\sigma_{x}$, $Y\overset{\text{def}}{=}i_{\mathbb{C}}\sigma_{y}$, $Z%
\overset{\text{def}}{=}\sigma_{z}$ defined in terms of the standard Pauli
operators $\sigma_{x}$, $\sigma_{y}$, $\sigma_{z}$ realizing the $\mathfrak{%
su}(2)$ algebra (throughout the paper $i_{\mathbb{C}}$ denotes the imaginary
unit of $\mathbb{C}$). They are such that $\{|0\rangle,|1\rangle\}$ are
eigenstates of $Z$, 
\begin{equation*}
XZ=-ZX \text{,}
\end{equation*}
and $Y=XZ$. The Pauli operators so defined suffice to describe all possible
errors occurring on a single qubit. Together with the identity operator $I$
(with $I=I_{2\times2}$) they form a multiplicative group if we allow them to
be multiplied by $-1$, i.e. $\{\pm I, \pm X, \pm Y, \pm Z\}$. We refer to
this group as the Pauli group $\mathcal{P}_{\mathcal{H}_{2}}$. Actually it
is a subgroup of the Pauli group realized through the standard sigma Pauli
operators and it coincides with the discrete version of the Heisenberg-Weyl
group.

For $n$-qubit errors we can then consider the Pauli group $\mathcal{P}_{%
\mathcal{H}_{2}^{\otimes n}}$ whose elements result from $n$-fold direct
products (see also \cite{gaitan}) 
\begin{equation*}
e(\lambda ,j_{1},\ldots ,j_{n})=(-)^{\lambda }e_{1}(j_{1})\otimes \text{...}%
\otimes e_{n}(j_{n})\text{.}
\end{equation*}%
The subscripts on the RHS label the qubits $1$,..., $n$, while $%
j_{k}=0,1,2,3 $, label respectively the operators $I_{k}$, $X_{k}$, $Y_{k}$, 
$Z_{k}$ acting on the $k$-th qubit. Furthermore $\lambda \in \left\{ 0\text{%
, }1\right\} $.

Since $e_{k}(2)=X_{k}Z_{k}$, the elements $e(\lambda ,j_{1},\ldots ,j_{n})$
can be rewritten as 
\begin{equation*}
e(\lambda ,a,b)=(-)^{\lambda }X{\left( a\right) }Z{\left( b\right) }\text{,}
\end{equation*}%
where $a=a_{1}$...$a_{n}$ and $b=b_{1}$...$b_{n}$ are bit strings of length $%
n$ and,%
\begin{equation*}
X\left( a\right) \overset{\text{def}}{=}\left( X_{1}\right) ^{a_{1}}\otimes 
\text{...}\otimes \left( X_{n}\right) ^{a_{n}}\text{, }Z\left( b\right) 
\overset{\text{def}}{=}\left( Z_{1}\right) ^{b_{1}}\otimes \text{...}\otimes
\left( Z_{n}\right) ^{b_{n}}\text{.}
\end{equation*}%
Observe that the order of the group $\mathcal{P}_{\mathcal{H}_{2}^{\otimes
n}}$ of $n$-qubits errors is $\left\vert \mathcal{P}_{\mathcal{H}%
_{2}^{\otimes n}}\right\vert =2^{2n+1}$. Since the factor $\pm $ in front of
an error makes no relevant difference on its action, we can actually assume
to work with the quotient group $\mathcal{P}_{\mathcal{H}_{2}^{\otimes
n}}/\left\{ \pm I\right\} $, with the major exception being the
determination whether elements commute or anti-commute.

The order of the quotient group is $\left\vert \mathcal{P}_{\mathcal{H}%
_{2}^{\otimes n}}/\left\{ \pm I\right\} \right\vert =2^{2n}$. Then, there is
a \emph{one-to-one} correspondence between $\mathcal{P}_{\mathcal{H}%
_{2}^{\otimes n}}/\left\{ \pm I\right\} $ and the $2n$-dimensional binary
vector space $\mathbb{F}_{2}^{2n}$ whose elements are bit strings of length $%
2n$ \cite{calderbank98}. A vector $v\in\mathbb{F}_{2}^{2n}$ is denoted $%
v=\left( a|b\right) $, where $a=a_{1}$...$a_{n}$ and $b=b_{1}$...$b_{n}$ are
bit strings of length $n$. Scalars take values in the Galois field $\mathbb{F%
}_{2}=\left\{ 0\text{, }1\right\} $ and vector addition adds components
modulo $2$. In short, we have the following correspondence $e(\lambda,a,b)
\in\mathcal{P}_{\mathcal{H}_{2}^{\otimes n}}\leftrightarrow v_{e}=\left(
a|b\right) \in\mathbb{F}_{2}^{2n}$.

A \emph{quantum stabilizer code} $\mathcal{C}$ is a vector space $\mathcal{C}%
\subseteq\mathcal{H}_{2}^{\otimes n}$ stabilized by an Abelian subgroup $%
\mathcal{S}\subseteq\mathcal{P}_{\mathcal{H}_{2}^{\otimes n}}$, i.e. such
that $\mathcal{S}|\psi\rangle=|\psi\rangle,\forall|\psi\rangle \in\mathcal{C}
$.

A quantum stabilizer code $C$ with stabilizer generators $g^{(1)}$,..., $%
g^{(n-1)}$ is a $2$-dimensional code space, i.e. a space where to encode a
single qubit. The codewords $|0\rangle _{L}$, $|1\rangle _{L}$ (basis
vectors for such code space) can be found as orthogonal eigenvectors
(corresponding to the eigenvalue $+1$) of any of the generators $g^{(j)}$.

The encoding operation then reads 
\begin{equation*}
\mathcal{H}_{2}\ni \left\vert 0\right\rangle \mapsto \left\vert
0_{L}\right\rangle \in \mathcal{H}_{2}^{\otimes n}\text{ and, }\mathcal{H}%
_{2}\ni \left\vert 1\right\rangle \mapsto \left\vert 1_{L}\right\rangle \in 
\mathcal{H}_{2}^{\otimes n}\text{.}
\end{equation*}%
In view of the correspondence between the Pauli group and the vector space $%
\mathbb{F}_{2}^{2n}$, let $v^{(j)}=\left( a^{(j)}|b^{(j)}\right) $ be the
image of the generators $g^{(j)}$ in $\mathbb{F}_{2}^{2n}$ and let introduce
the so called \emph{parity check matrix} 
\begin{equation}
H\overset{\text{def}}{=}\left( 
\begin{array}{c}
\left( a^{(1)}|b^{(1)}\right) \\ 
\left( a^{(2)}|b^{(2)}\right) \\ 
\vdots \\ 
\left( a^{(n-1)}|b^{(n-1)}\right)%
\end{array}%
\right) \,.  \label{pcm}
\end{equation}%
Then, for an error $e\in \mathcal{P}_{\mathcal{H}_{2}^{\otimes
n}}\leftrightarrow v_{e}=\left( a|b\right) \in \mathbb{F}_{2}^{2n}$, the
error syndrome $S(e)$ is given by the bit string \cite{gaitan} 
\begin{equation}
S(e)=Hv_{e}=l_{1}\text{...}l_{n-1}\text{,}  \label{syndrom}
\end{equation}%
where $l_{j}=H^{T}\left( j\right) \cdot v_{e}$.

Errors with non-vanishing error syndrome are \emph{detectable}. They
correspond to operators not in $\mathcal{S}$ and not commuting with those in 
$\mathcal{S}$. That is, a set of error operators $\mathcal{E}\subseteq 
\mathcal{P}_{\mathcal{H}_{2}^{\otimes n}}$ is detectable if $\mathcal{E}%
\notin\mathcal{Z}(\mathcal{S})-\mathcal{S}$, with $\mathcal{Z}(\mathcal{S})$
the centralizer of the subgroup $\mathcal{S}$. Furthermore, the set of error
operators $\mathcal{E}\subseteq\mathcal{P}_{\mathcal{H}_{2}^{\otimes n}}$ is 
\emph{correctable} if the set given by $\mathcal{E}^{\dagger}\mathcal{E}$ is
in turn detectable \cite{knill02}, i.e. $\mathcal{E}^{\dag}\mathcal{E}\notin%
\mathcal{Z}(\mathcal{S})-\mathcal{S}$.

It would be awfully tedious to identify either detectable errors or sets of
correctable errors. However, the quantum stabilizer formalism allows to
simplify such task \cite{daniel-thesis}. This is a consequence of the fact
that by means of such formalism it is sufficient to study the effect of the
error operators on the generators of the stabilizer and not on the codewords
themselves. Actually the syndrome extraction corresponds to measure the
stabilizer generators.

Finally, we denote by $\left[ \left[ n\text{, }k\text{, }d_{\mathcal{C}}%
\right] \right] $\ a quantum stabilizer code $\mathcal{C}$\ with code
parameters $n$ (the length), $k$ (the dimension) and $d_{\mathcal{C}}$ (the
distance) encoding $k$-logical qubits into $n$-physical qubits and
correcting $\lfloor \frac{d_{\mathcal{C}}-1}{2}\rfloor $-qubit errors ( $%
\lfloor x\rfloor $ denotes the largest integer less than $x$).

%%%%%%%%%%%%%%%%%%%%%%%%%%%%%%%%%%%%%%%%

\subsection{Generalized Pauli group and qudit-encoding}

A qudit is a $d$-dimensional quantum system with associated complex Hilbert
space $\mathcal{H}_{d}$. On this space we can introduce a generalized
version of the Pauli operators $X$ and $Z$ considered in the previous
Subsection. They can be defined through their action on the canonical basis $%
\{|k\rangle \}_{k\in\mathbb{Z}_{d}}$ of $\mathcal{H}_{d}$ \cite{daniel-CSF}, 
\begin{equation}
X\left\vert k\right\rangle =\left\vert k\oplus1\right\rangle \text{ and, }%
Z\left\vert k\right\rangle =\omega^{k}\left\vert k\right\rangle \text{,}%
\quad k\in\mathbb{Z}_{d},  \label{defXZ}
\end{equation}
where "$\oplus$" denotes addition of integers modulo $d$ and $\omega \overset%
{\text{def}}{=}\exp\left( i_{\mathbb{C}}\frac{2\pi}{d}\right) $ is a
primitive $d$th root of unity ($\omega^{d}=1$). The $X$ and $Z$ operators so
defined are unitary ($X^{\dag}=X^{-1}$ and $Z^{\dag}=Z^{-1}$), but not
Hermitian, and satisfy $X^{d}=Z^{d}=I$ (with $I=I_{d\times d}$) together
with the commutation relations 
\begin{equation}
X^{a}Z^{b}=\omega^{-ab}Z^{b}X^{a}\text{.}  \label{CRPaulid}
\end{equation}
It is then possible to consider the Pauli group $\mathcal{P}_{\mathcal{H}%
_{d}}$ consisting of all operators $e$ of the form%
\begin{equation*}
e\left( l\text{, }n\text{, }m\right) =\omega^{l}X^{n}Z^{m}\text{,}
\end{equation*}
where $l,n,m\in\mathbb{Z}_{d}$. Similarly to the previous Subsection, for
errors on a qudit we may refer to the quotient group $\mathcal{P}_{\mathcal{H%
}_{d}}/\{\omega^{l}I|l=0,\ldots,d-1\}$.

In addition to the reasons stated in the introductory Section about passing
from block to embedding codes, another motivation is to understand whether
or not finite dimensional versions of the shift-resistant quantum codes of
Ref.\cite{gotty-preskill} are effective. Hence, following Ref. \cite%
{gotty-preskill}, we consider $d=2r_{1}r_{2}$, and introduce a code $%
\mathcal{C}$ stabilized by the Abelian subgroup $\mathcal{S}\subset \mathcal{%
P}_{\mathcal{H}_{d}}$ generated by $X^{2r_{1}}$, $Z^{2r_{2}}$. The codewords
(basis for $\mathcal{C}$) are eigenstates of $Z^{2r_{2}}$ and $X^{2r_{1}}$
with eigenvalue one. Hence they only contain $|k\rangle $s with values of $k$
that are multiples of $r_{1}$ and that are invariant under a shift by $%
2r_{1} $. They read 
\begin{align}
& |0_{L}\rangle ={\frac{1}{\sqrt{r_{2}}}}\left( |0\rangle +|2r_{1}\rangle
+\dots +|2(r_{2}-1)r_{1}\rangle \right) \,\text{,}  \label{codeword0} \\
& |1_{L}\rangle ={\frac{1}{\sqrt{r_{2}}}}\left( |r_{1}\rangle +\dots
+|(2(r_{2}-1)+1)r_{1}\rangle \right) \,\text{.}  \label{codeword1}
\end{align}

The encoding operation becomes in this case 
\begin{equation*}
\mathcal{H}_{2}\ni\left\vert 0\right\rangle \mapsto\left\vert
0_{L}\right\rangle \in\mathcal{H}_{d}\text{ and, }\mathcal{H}%
_{2}\ni\left\vert 1\right\rangle \mapsto\left\vert 1_{L}\right\rangle \in%
\mathcal{H}_{d}\text{.}
\end{equation*}
If the states \eqref{codeword0}, \eqref{codeword1} undergo an amplitude
shift, the value of $k$ modulo $r_{1}$ is determined by measuring the
stabilizer generator $Z^{2r_{2}}$, and the shift can be corrected by
adjusting $k$ to the nearest multiple of $r_{1}$.

The codewords in the basis of $X$'s eigenstates can be found by observing
that the eigenstates of $X$ and $Z$ operators are connected by the Fourier
transform 
\begin{equation*}
\widetilde{\left\vert i\right\rangle }=\frac{1}{\sqrt{d}}\sum_{j=0}^{d-1}%
\omega^{-ij}\left\vert j\right\rangle ~,
\end{equation*}
where $X\widetilde{\left\vert i\right\rangle }=\omega^{i}\widetilde{%
\left\vert i\right\rangle }$. Then, it turns out that 
\begin{align}
& |0_{L}\rangle= {\frac{1}{\sqrt{r_{1}}}}\left( \widetilde{|0\rangle} +%
\widetilde{|2r_{2}\rangle} +\dots+\widetilde{|2(r_{1}-1)r_{2}\rangle}\right)
\,,  \label{codeword0FT} \\
& |1_{L}\rangle= {\frac{1}{\sqrt{r_{1}}}}\left( \widetilde{|r_{2}\rangle} +
\dots+\widetilde{|(2(r_{1}-1)+1)r_{2}\rangle}\right) \,.  \label{codeword1FT}
\end{align}
The codewords \eqref{codeword0FT}, \eqref{codeword1FT}, have the same form
of \eqref{codeword0}, \eqref{codeword1}, but with $r_{1}$ and $r_{2}$
interchanged. Hence, if they undergo a phase shift, the value of $k$ modulo $%
r_{2}$ is determined by measuring the stabilizer generator $X^{2r_{1}}$, and
the shift can be corrected by adjusting $k$ to the nearest multiple of $%
r_{2} $.

To understand what is the set of correctable errors according to the
condition $\mathcal{E}^{\dag}\mathcal{E}\notin\mathcal{Z}(\mathcal{S})-%
\mathcal{S}$ we have with the help of \eqref{CRPaulid} 
\begin{align}
\left( X^{a^{\prime}}Z^{b^{\prime}}\right) ^{\dagger}(X^{a}Z^{b})X^{2r_{1}}
& =e^{2\pi i(b-b^{\prime})/r_{2}}~X^{2r_{1}}\left(
X^{a^{\prime}}Z^{b^{\prime}}\right) ^{\dag}(X^{a}Z^{b})\,,  \notag \\
\left( X^{a^{\prime}}Z^{b^{\prime}}\right) ^{\dagger}(X^{a}Z^{b})Z^{2r_{2}}
& =e^{-2\pi i(a-a^{\prime})/r_{1}}~Z^{2r_{2}}\left(
X^{a^{\prime}}Z^{b^{\prime}}\right) ^{\dag}(X^{a}Z^{b})\,.  \notag
\end{align}
The phases on the right hand sides are non-trivial only if 
\begin{equation}
|a-a^{\prime}|<\frac{r_{1}}{2}\,,\quad{\text{and}}\quad|b-b^{\prime}|<\frac{%
r_{2}}{2}\,.  \label{oggi}
\end{equation}
Therefore, the code $\mathcal{C}$ can correct all shifts corresponding to
these conditions. They amount to $r_{1}r_{2}=d/2$ and this is also the
number of possible error syndromes.

Finally, notice that other families of qudit codes could be constructed by
generalizing the Pauli operators in a different way, e.g. by making them
Hermitian.

%%%%%%%%%%%%%%%%%%%%%%%%%%%%%%%%%%%%%%%%%

\section{The Weyl quantum channel as error model}

In this Section, we first discuss the Weyl quantum channel for qudit states
and for qubit states, and then the entanglement fidelity as quantifier of
codes performances.

\subsection{General form of the Weyl quantum channel}

Consider a completely positive trace preserving map (CPT map or quantum
channel) $\Lambda ^{(d)}$, 
\begin{equation*}
\Lambda ^{(d)}:\mathfrak{S}\left( \mathcal{H}_{d}\right) \ni \rho
\longmapsto \Lambda ^{(d)}\left( \rho \right) \in \mathfrak{S}\left( 
\mathcal{H}_{d}\right) \text{,}
\end{equation*}%
where $\mathfrak{S}\left( \mathcal{H}_{d}\right) $ is the set of positive
unit trace linear operators in $\mathcal{H}_{d}$. Then, $\Lambda ^{(d)}$ is
called bistochastic if, 
\begin{equation*}
\Lambda ^{(d)}\left( \frac{1}{d}I\right) \overset{\text{def}}{=}\frac{1}{d}I%
\text{,}
\end{equation*}%
where $I$ ($=I_{d\times d}$) is the identity operator in $\mathcal{H}_{d}$.
The $d$-dimensional Weyl channel is a bistochastic quantum channel of the
following form \cite{amosov}, 
\begin{equation}
\Lambda _{\text{Weyl}}^{\left( d\right) }\left( \rho \right) \overset{\text{
def}}{=}\sum_{n\text{, }m=0}^{d-1}\pi ({n\text{, }m})\,U_{n\text{, }m}\rho
U_{n\text{, }m}^{\dagger }\text{,}  \label{dweyl}
\end{equation}%
where $\rho \in \mathfrak{S}\left( \mathcal{H}_{d}\right) $ and $\pi ({n%
\text{, }m})$ is an arbitrary probability distribution, thus respecting $%
0\leq \pi ({n\text{, }m})\leq 1$, $\sum_{n\text{, }m=0}^{d-1}\pi ({n\text{, }%
m})=1$. The unitary Weyl operators $U_{n\text{, }m}$ in (\ref{dweyl}) are
defined as, 
\begin{equation*}
U_{n\text{, }m}\overset{\text{def}}{=}\sum_{k=0}^{d-1}\exp \left( i_{\mathbb{%
C}}\frac{2\pi }{d}km\right) \left\vert k\oplus n\right\rangle \left\langle
k\right\vert \text{,}
\end{equation*}%
where "$\oplus $", as specified earlier, denotes the addition of integers
modulo $d$. They also satisfy the (Weyl) commutation relations %
\eqref{CRPaulid}, i.e. 
\begin{equation*}
U_{n\text{, }m}U_{n^{\prime }\text{, }m^{\prime }}=\exp \left[ i_{\mathbb{C}}%
\frac{2\pi }{d}\left( n^{\prime }m-nm^{\prime }\right) \right] U_{n^{\prime }%
\text{, }m^{\prime }}U_{n\text{, }m}\text{,}
\end{equation*}%
where $0\leq n$, $n^{\prime }$, $m$, $m^{\prime }\leq d-1$. Notice that $U_{n%
\text{, }m}$ may be rewritten as $U_{n\text{, }m}=U_{n\text{, }0}U_{0\text{, 
}m}$. Furthermore, $U_{n,0}\equiv X^{n}$ and $U_{0,m}\equiv Z^{m}$%
. Therefore in what follows we will consider the $d$-dimensional Weyl
channel acting as 
\begin{equation}
\Lambda _{\text{Weyl}}^{\left( d\right) }\left( \rho \right) \overset{\text{%
def}}{=}\sum_{n\text{, }m=0}^{d-1}\pi ({n\text{, }m})X^{n}Z^{m}\rho \left(
X^{n}Z^{m}\right) ^{\dagger }\text{.}  \label{Wch}
\end{equation}%
For $d=2$ we have the most general channel acting on a qubit 
\begin{equation}
\Lambda _{\text{Weyl}}^{\left( 2\right) }\left( \rho \right) =\pi ({0,0}%
)\rho +\pi ({1,0})X\rho X+\pi ({1,1})Y\rho Y+\pi ({0,1})Z\rho Z\text{.}
\label{Wch2}
\end{equation}%
For instance, if we take $\pi ({0,0})\overset{\text{def}}{=}1-p\text{, }\pi (%
{1,0})=\pi ({1,1})=\pi ({0,1})\overset{\text{def}}{=}\frac{p}{3}$, the
channel $\Lambda _{\text{Weyl}}^{\left( 2\right) }$ becomes the standard
symmetric qubit depolarizing channel.

To justify the choice of Weyl's error model, we point out that the quantum
codes employed here are designed to error-correct arbitrary quantum errors
such as $X$-errors, $Z$-errors and combinations of the two ($Y$-errors).
Since the Kraus decomposition of the Weyl channel is defined in terms of
powers of these aforementioned error operators, it certainly constitutes
a natural test-bed where quantifying the performance of
the selected qudit codes.  
Such a test-bed turns also out to be very general, with the possibility of encompassing
physically relevant scenarios.

Hereafter, when we consider the two-dimensional Weyl channel, we mean to
take into consideration the following channel parametrization 
\begin{equation}
\pi\left( n\text{, }m\right) \overset{\text{def}}{=}\pi_{X}\left( n\right)
\pi_{Z}\left( m\right) \equiv\pi\left( n\right) \pi\left( m\right) \text{,}
\label{pparam}
\end{equation}
with, 
\begin{equation}
\pi\left( 1\right) =\pi\left( -1\right) \overset{\text{def}}{=}p\text{, }%
\pi\left( 0\right) \overset{\text{def}}{=}1-p\text{.}
\end{equation}
In this way we have 
\begin{equation}
\Lambda_{\text{Weyl}}^{\left( 2\right) }\left( \rho\right) =\left(
1-p\right) ^{2}\rho+(1-p)pX\rho X^{\dagger}+p^{2}Y\rho Y^{\dagger
}+p(1-p)Z\rho Z^{\dagger}\text{.}
\end{equation}
Then, we can also consider the possibility of having asymmetric $X$ and $Z$
errors' probabilities \cite{carlo-PRA, ioffe07, evans07, stephens08,
sarvepalli08, aly08}. In such a case the probabilities $\pi\left( n\text{, }%
m\right) $ are defined as follows, 
\begin{equation}
\pi^{\left( \text{asymmetric}\right) }\left( n\text{, }m\right) \overset{%
\text{def}}{=}\pi_{X}\left( n\right) \pi_{Z}\left( m\right) \text{,}
\label{piasym}
\end{equation}
with $\pi_{X}$ and $\pi_{Z}$ not identical. In particular, we shall consider 
\begin{equation}
\pi_{X}(1)=\kappa\pi_{Z}(1),  \label{pik}
\end{equation}
with $\pi_{Z}\left( 1\right) \overset{\text{def}}{=}p$ and $\kappa\in\left[
0,1/p\right] $ to guarantee that $\kappa p$ is a valid probability value.
Notice that for $\kappa=1$ we recover the symmetric case.

Finally, when considering block codes on $n$ qubits, the error map simply
becomes $\Lambda^{(2)\otimes n}_{\text{Weyl}}$ acting on $\rho\in \mathfrak{S%
}(\mathcal{H}_{2}^{\otimes n})$.

\subsection{Entanglement Fidelity}

Entanglement fidelity is a reliable performance measure of the efficiency of
quantum error correcting codes \cite{carlo-OSID}. Suppose a two dimensional
code $\mathcal{C}$ is such that $\mathcal{C}\subset\mathcal{H}$ with $\text{%
dim}_{\mathbb{C}}\mathcal{H}=N$ (here $\mathcal{H}$ can be either $\mathcal{H%
}_{2}^{\otimes n}$ or $\mathcal{H}_{d}$, hence $N$ is either $2^{n}$ or $d$%
). Then consider errors taking place through a CPT map $\Lambda: \mathfrak{S}%
\left( \mathcal{H}\right) \rightarrow\mathfrak{S}\left( \mathcal{H}\right) $
(which can be either $\Lambda^{(2)\otimes n}$ or $\Lambda^{(d)}$) written,
in terms of the Kraus decomposition, as 
\begin{equation*}
\Lambda\left( \rho\right) =\sum_{k}A_{k} \rho A_{k}^{\dagger} \text{.}
\end{equation*}
To recover the errors by means of the code $\mathcal{C}$ a \emph{recovery}
operation must be applied according to the syndrome extraction. Suppose it
is described by a CPT map $\mathcal{R}:\mathfrak{S}\left( \mathcal{H}\right)
\rightarrow\mathfrak{S}\left( \mathcal{H}\right) $ 
\begin{equation*}
\mathcal{R}\left( \rho\right) =\sum_{k}R_{k} \rho R_{k}^{\dagger} \text{.}
\end{equation*}
If the code is effective, we expect that the map resulting from the
composition of $\Lambda$ and $\mathcal{R}$ restricted to the subspace $%
\mathcal{C}$, namely $\left[ \mathcal{R}\circ\Lambda\right] _{\vert _{%
\mathcal{C}}}$, will be close to the identity map $\mathrm{{id}_{\mathcal{C}}%
}$ on $\mathcal{C}$. In order to evaluate this closeness we can consider a
state $\rho=\text{tr}_{\mathcal{C}}\left\vert \psi\right\rangle \left\langle
\psi\right\vert $ written in terms of a purification $\left\vert
\psi\right\rangle \in\mathcal{C}\otimes\mathcal{C}$ and see how well
entanglement (between $\mathcal{C}$ and the reference system identical to $%
\mathcal{C}$) is preserved by means of 
\begin{equation*}
\mathcal{F}\left( \rho\text{, }\left[ \mathcal{R}\circ\Lambda\right]
_{\vert_{\mathcal{C}}}\right) \overset{\text{def}}{=} \left\langle \psi%
\Big\vert \left( \left[ \mathcal{R}\circ\Lambda^{(N)}\right] _{\vert_{%
\mathcal{C}}} \otimes\text{id}_{\mathcal{C}}\right) \left( \left\vert
\psi\right\rangle \left\langle \psi\right\vert \right) \Big\vert%
\psi\right\rangle \text{,}
\end{equation*}
This is the entanglement fidelity \cite{schumy} for the map $\left[ \mathcal{%
R}\circ\Lambda\right] _{\vert_{\mathcal{C}}}$.

In terms of the Kraus error operators, $\mathcal{F}$ can be rewritten as 
\cite{mike} 
\begin{equation*}
\mathcal{F}\left( \rho\text{, }\left[ \mathcal{R}\circ\Lambda\right] _{|_{%
\mathcal{C}}}\right) =\sum_{j,k}\left\vert \text{tr}\left[ R_{j}A_{k}\right]
_{|_{\mathcal{C}}}\right\vert ^{2}\text{.}
\end{equation*}
Finally, choosing a purification described by a maximally entangled unit
vector $\left\vert \psi\right\rangle \in\mathcal{C}\otimes\mathcal{C}$ for
the mixed state $\rho=\frac{1}{\text{dim}_{\mathbb{C}}\mathcal{C}}I_{%
\mathcal{C}}$, we obtain 
\begin{equation}
\mathcal{F}\left( \frac{1}{2}I_{\mathcal{C}}\text{, }\left[ \mathcal{R}%
\circ\Lambda\right] _{|_{\mathcal{C}}}\right) =\frac{1}{2^{2}}\sum
_{j,k}\left\vert \text{tr}\left[ R_{j}A_{k}\right] _{|_{\mathcal{C}%
}}\right\vert ^{2}\text{.}  \label{nfi}
\end{equation}
This is the expression we will use in the following.

%%%%%%%%%%%%%%%%%%%%%%%%%%%%%%%%%%%%%%%%%%%%

\section{Qudit codes for Weyl errors}

In this Section, we analyze in details how the qudit codes devised in Sec.
II work on the Weyl channel for $d=18$ and $d=50$ and determine their
performance by means of the entanglement fidelity.

Our main motivation to use the qudit codes with $d=18$\ and $d=50$\ is that
they represent the lowest-dimensional perfect qudit systems where a
two-dimensional quantum systems (a qubit) can be encoded and protected
against arbitrary shift errors of the form $X^{n}Z^{m}$ \ by one and two
units, respectively. Furthermore, the physical dimensionality of these codes
is chosen so to be as much as possible comparable with the physical
dimensionality of code-spaces characterizing well-known standard stabilizer
error correction schemes capable of correcting arbitrary single-qubit errors
and, possibly, few two-qubits errors. For this reason, the five \cite%
{laflamme, bennett} and seven-qubit \cite{steane, calderbank} quantum
stabilizer codes seem to be a convenient choice. In particular, recalling
that a stabilizer code is perfect if all the eigenvalues of the generators
constitute valid syndromes for correcting an error, it turns out that both
the five and the qudit code with $d=18$\ are perfect and require minimal
quantum resources for their task.

\subsection{The $d=18$ qudit code}

\emph{Encoding}. The encoding operation is characterized by, 
\begin{equation*}
\mathcal{H}_{2}\ni\left\vert 0\right\rangle \mapsto\left\vert
0_{L}\right\rangle \in\mathcal{H}_{18}\text{ and, }\mathcal{H}
_{2}\ni\left\vert 1\right\rangle \mapsto\left\vert 1_{L}\right\rangle \in%
\mathcal{H}_{18}\text{,}
\end{equation*}
where the codewords $\left\vert 0_{L}\right\rangle $ and $\left\vert
1_{L}\right\rangle $ are defined according to \eqref{codeword0}, %
\eqref{codeword1} as,%
\begin{equation*}
\left\vert 0_{L}\right\rangle \overset{\text{def}}{=}\frac{1}{\sqrt{3}}\left[
\left\vert 0\right\rangle +\left\vert 6\right\rangle +\left\vert
12\right\rangle \right] \text{ and, }\left\vert 1_{L}\right\rangle \overset{%
\text{def}}{=}\frac{1}{\sqrt{3}}\left[ \left\vert 3\right\rangle +\left\vert
9\right\rangle +\left\vert 15\right\rangle \right] \text{,}
\label{encoding1}
\end{equation*}
respectively. The stabilizer group $\mathcal{S}$ of this code is generated
by the two error operators $X^{6}$ and $Z^{6}$. Here $r_{1}=r_{2}=3$.

A simple calculation shows that provided we restrict our focus to the
two-dimensional code space $\mathcal{C}^{\left( d=18\right) }\subset 
\mathcal{H}_{18}$, the following identities hold,%
\begin{align}
I& =X^{6}=X^{12}\text{, }X=X^{7}=X^{13}\text{, }X^{2}=X^{8}=X^{14}\text{, }%
X^{3}\equiv X^{-3}=X^{9}=X^{15}\text{,}  \notag \\
&  \notag \\
\text{ }X^{4}& \equiv X^{-2}=X^{10}=X^{16}\text{, }X^{5}\equiv
X^{-1}=X^{11}=X^{17}\text{,}  \notag
\end{align}%
and,%
\begin{align}
I& =Z^{6}=Z^{12}\text{, }Z=Z^{7}=Z^{13}\text{, }Z^{2}=Z^{8}=Z^{14}\text{, }%
Z^{3}\equiv Z^{-3}=Z^{9}=Z^{15}\text{, }  \notag \\
&  \notag \\
Z^{4}& \equiv Z^{-2}=Z^{10}=Z^{16}\text{, }Z^{5}\equiv Z^{-1}=Z^{11}=Z^{17}%
\text{.}  \notag
\end{align}%
Therefore the total number of error operators to be considered amounts to be 
$36$ rather than $18^{2}$. Then, the Weyl channel \eqref{Wch} may be
rewritten as, 
\begin{equation*}
\Lambda ^{\left( d=18\right) }\left( \rho \right) =\sum_{k=0}^{35}A_{k}\rho
A_{k}^{\dagger }\text{,}
\end{equation*}%
where we have relabeled the error operators as follows, 
\begin{align}
& A_{0}\overset{\text{def}}{=}\sqrt{\pi \left( 0\text{, }0\right) }I\text{, }%
A_{1}\overset{\text{def}}{=}\sqrt{\pi \left( 0\text{, }1\right) }Z,\ldots
,A_{18}\overset{\text{def}}{=}\sqrt{\pi \left( -1\text{, }3\right) }%
X^{-1}Z^{3},  \notag \\
&  \notag \\
& \ldots ,A_{34}\overset{\text{def}}{=}\sqrt{\pi \left( 3,-2\right) }%
X^{3}Z^{-2}\text{, }A_{35}\overset{\text{def}}{=}\sqrt{\pi \left( 3\text{, }%
3\right) }X^{3}Z^{3}\text{.}  \notag
\end{align}%
In the \emph{symmetric case}, the probabilities $\pi \left( n\text{, }%
m\right) $ are defined like in Eq.\eqref{pparam}, where now 
\begin{equation*}
\pi \left( 1\right) =\pi \left( -1\right) \overset{\text{def}}{=}p\text{, }%
\pi \left( 2\right) =\pi \left( -2\right) \overset{\text{def}}{=}p^{2}\text{%
, }\pi \left( 3\right) \equiv \pi \left( -3\right) \overset{\text{ def}}{=}%
p^{3}\text{, }\pi \left( 0\right) \overset{\text{def}}{=}1-2p-2p^{2}-p^{3}%
\text{.}
\end{equation*}

\emph{Correctability}. According to (\ref{oggi}) with $r_{1}=r_{2}=3$ the
set of correctable errors $\mathcal{A}_{\text{correctable}}^{\left(
d=18\right) }$ is given by the following $9$ errors, 
\begin{equation}
\mathcal{A}_{\text{correctable}}^{\left( d=18\right) }=\left\{ A_{0}\text{, }%
A_{1}\text{, }A_{2}\text{, }A_{6}\text{, }A_{7}\text{, }A_{8}\text{, }A_{12}%
\text{, }A_{13}\text{, }A_{14}\right\} \text{,}  \label{18correctable}
\end{equation}
where, 
\begin{align}
& A_{0}\overset{\text{def}}{=}\sqrt{\pi\left( 0\text{, }0\right) }I\text{, }%
A_{1}\overset{\text{def}}{=}\sqrt{\pi\left( 0\text{, }1\right) }Z\text{, }%
A_{2}\overset{\text{def}}{=}\sqrt{\pi\left( 0\text{, }-1\right) }Z^{-1}\text{%
, }A_{6}\overset{\text{def}}{=}\sqrt{\pi\left( 1\text{, }0\right) }X\text{, }%
A_{7}\overset{\text{def}}{=}\sqrt{\pi\left( 1\text{, }1\right) }XZ\text{, } 
\notag \\
&  \notag \\
& A_{8}\overset{\text{def}}{=}\sqrt{\pi\left( 1\text{, }-1\right) }XZ^{-1}%
\text{, }A_{12}\overset{\text{def}}{=}\sqrt{\pi\left( -1\text{, }0\right) }%
X^{-1}\text{, }A_{13}\overset{\text{def}}{=}\sqrt{\pi\left( -1\text{, }%
1\right) }X^{-1}Z\text{, }A_{14}\overset{\text{def}}{=}\sqrt {\pi\left( -1%
\text{, }-1\right) }X^{-1}Z^{-1}\text{.}  \notag
\end{align}

\emph{Recovery Operators}. Following the recipe provided in \cite{KL}, it
turns out that the two \ $9$-dimensional orthogonal subspaces $\mathcal{V}%
^{0_{L}}$ and $\mathcal{V}^{1_{L}}$ of $\mathcal{H}_{18}$ generated by the
action of $\mathcal{A}_{\text{correctable}}^{\left( d=18\right) }$ on $%
\left\vert 0_{L}\right\rangle $ and $\left\vert 1_{L}\right\rangle $ are
given by, 
\begin{equation*}
\mathcal{V}^{0_{L}}=Span\left\{ \left\vert v_{k}^{0_{L}}\right\rangle =\frac{%
A_{k}}{\sqrt{\pi_{k}}}\left\vert 0_{L}\right\rangle \text{, }\right\} \text{,%
}\quad\text{and}\quad\mathcal{V}^{1_{L}}=Span\left\{ \left\vert
v_{k}^{1_{L}}\right\rangle =\frac{A_{k}}{\sqrt{\pi_{k}}}\left\vert
1_{L}\right\rangle \right\} \text{,}
\end{equation*}
with $k\in\mathcal{I}^{\left( d=18\right) }\overset{\text{def}}{=}\left\{ 0%
\text{, }1\text{, }2\text{, }6\text{, }7\text{, }8\text{, }12\text{, }13%
\text{, }14\right\} $. The coefficients $\sqrt{\pi_{k}}$ denote the errors
amplitudes where, for instance, $\sqrt{\pi_{8}}\overset{\text{def}}{=}\sqrt{%
\pi\left( 1\text{, }-1\right) }$. Notice that $\left\langle
v_{k}^{i_{L}}|v_{k^{\prime}}^{j_{L}}\right\rangle
=\delta_{kk^{\prime}}\delta_{ij}$ with $k$, $k^{\prime}\in\mathcal{I}%
^{\left( d=18\right) }$ and $i$, $j\in\left\{ 0\text{, }1\right\} $.
Therefore, it follows that $\mathcal{V}^{0_{L}}\oplus\mathcal{V}^{1_{L}}=%
\mathcal{H}_{18}$. The recovery superoperator $\mathcal{R}%
\leftrightarrow\left\{ R_{k}\right\} $ with $k\in\mathcal{I}^{\left(
d=18\right) }$ is defined by means of \cite{KL}, 
\begin{equation*}
R_{k}=\left\vert 0_{L}\right\rangle \left\langle v_{k}^{0_{L}}\right\vert
+\left\vert 1_{L}\right\rangle \left\langle v_{k}^{1_{L}}\right\vert \text{.}
\label{may1}
\end{equation*}
In the case under investigation, the entanglement fidelity \eqref{nfi}
reads, 
\begin{equation*}
\mathcal{F}^{\left( d=18\right) }\left( p\right) \overset{\text{def}}{=}%
\mathcal{F}^{\left( d=18\right) }\left( \frac{1}{2}I_{2\times2}\text{, }%
\mathcal{R}\circ\Lambda^{\left( d=18\right) }\right) =\frac{1}{\left(
2\right) ^{2}}\sum_{l=0}^{35}\sum\limits_{k\in\mathcal{I}^{\left(
d=18\right) }}\left\vert \text{tr}\left( \left[ R_{k}A_{l}\right] _{|%
\mathcal{C}^{\left( d=18\right) }}\right) \right\vert ^{2}\text{,}
\label{f18}
\end{equation*}
where, 
\begin{equation*}
\left[ R_{k}A_{l}\right] _{|\mathcal{C}^{\left( d=18\right) }}\overset{\text{%
def}}{=}\left( 
\begin{array}{cc}
\left\langle 0_{L}|R_{k}A_{l}|0_{L}\right\rangle & \left\langle
0_{L}|R_{k}A_{l}|1_{L}\right\rangle \\ 
\left\langle 1_{L}|R_{k}A_{l}|0_{L}\right\rangle & \left\langle
1_{L}|R_{k}A_{l}|1_{L}\right\rangle%
\end{array}
\right) \text{.}
\end{equation*}
After a simple calculation, $\mathcal{F}^{\left( d=18\right) }\left(
p\right) $ becomes, 
\begin{equation}
\mathcal{F}^{\left( d=18\right) }\left( p\right)
=1-4p^{2}-2p^{3}+4p^{4}+4p^{5}+p^{6}\text{.}  \label{f1}
\end{equation}
We stress that this error correction scheme is effective as long as the
failure probability $1-\mathcal{F}^{\left( d=18\right) }\left( p\right) $ is
strictly smaller than the error probability $p$. This implies that the $d=18$
-dimensional qudit code is effective only when $0\leq p\lesssim0.24$.
Furthermore, we point out that in its range of effectiveness, $\mathcal{F}%
^{\left( d=18\right) }\left( p\right) $ in (\ref{f1}) is a monotonic
decreasing function of $p$.

\subsubsection{Asymmetric errors}

By repeating the steps of the previous Subsection using Eqs.\eqref{piasym}
and \eqref{pik} we get 
\begin{equation}
\mathcal{F}_{\text{asymmetric}}^{(d=18)}\left( p\right)
=1-2p^{2}-p^{3}+\kappa^{2}\left( -2p^{2}+4p^{4}+2p^{5}\right)
+\kappa^{3}\left( -p^{3}+2p^{5}+p^{6}\right) \text{.}  \label{prima}
\end{equation}
Eqs.(\ref{f1}) and (\ref{prima}) become, to the leading order in $p$ with $%
p\ll1$, 
\begin{equation*}
\mathcal{F}^{(d=18)}\left( p\right) \overset{p\ll1}{\approx}1-4p^{2},\quad%
\mathcal{F}_{\text{asymmetric}}^{(d=18)}\left( p\right) \overset{p\ll1}{
\approx} 1-2(1+\kappa^{2})p^{2} \text{.}
\end{equation*}
It results that the presence of asymmetric Weyl errors with $\kappa<1$
increases the performance of the correction scheme. This can be understood
by noticing that as soon as $\kappa\to0$ the noise model becomes
classical-like and the errors become of a single type, namely $Z$ type,
hence more easy to correct (this limiting case is similar to that
investigated in Ref.\cite{pirandola}). On the contrary, for $\kappa>1$ the
performance of the code is lowered by errors asymmetries.

%%%%%%%%%%%%%%%%%%%%%%%%%%%%%%%

\subsection{The $d=50$ qudit code}

The encoding operation in this case is characterized by, 
\begin{equation*}
\mathcal{H}_{2}\ni\left\vert 0\right\rangle \mapsto\left\vert
0_{L}\right\rangle \in\mathcal{H}_{50}\text{ and, }\mathcal{H}
_{2}\ni\left\vert 1\right\rangle \mapsto\left\vert 1_{L}\right\rangle \in%
\mathcal{H}_{50}\text{,}
\end{equation*}
where the codewords $\left\vert 0_{L}\right\rangle $ and $\left\vert
1_{L}\right\rangle $ are defined according to \eqref{codeword0}, %
\eqref{codeword1} as, 
\begin{equation*}
\left\vert 0_{L}\right\rangle \overset{\text{def}}{=}\frac{1}{\sqrt{5}}\left[
\left\vert 0\right\rangle +\left\vert 10\right\rangle +\left\vert
20\right\rangle +\left\vert 30\right\rangle +\left\vert 40\right\rangle %
\right] \text{ and, }\left\vert 1_{L}\right\rangle \overset{\text{def}}{=} 
\frac{1}{\sqrt{5}}\left[ \left\vert 5\right\rangle +\left\vert
15\right\rangle +\left\vert 25\right\rangle +\left\vert 35\right\rangle
+\left\vert 45\right\rangle \right] \text{,}  \label{encoding2}
\end{equation*}
respectively. The stabilizer group $\mathcal{S}$ of this code is generated
by the two error operators $X^{10}$ and $Z^{10}$. Here $r_{1}=r_{2}=5$.

Indeed, it can be shown that provided we restrict our focus to the
two-dimensional code space $C^{\left( d=50\right) }\subset H_{50}$, the set
of all ($50^{2}$) non-normalized errors can be reduced to $100$\ operators.
Following the same line of reasoning of the previous Subsection, it is
possible to find the recovery superoperator. After some algebraic
calculations, we arrive at the following expression for the entanglement
fidelity 
\begin{equation}
\mathcal{F}^{\left( d=50\right) }\left( p\right)
=1-4p^{3}-4p^{4}-2p^{5}+4p^{6}+8p^{7}+8p^{8}+4p^{9}+p^{10}\text{.}
\label{f3}
\end{equation}%
We emphasize that this error correction scheme is effective as long as the
failure probability $1-\mathcal{F}^{\left( d=50\right) }\left( p\right) $ is
strictly smaller than the error probability $p$. This implies that the $d=50$
-dimensional qudit code is effective only when $0\leq p\lesssim 0.43$. This $%
p$-range of effectiveness is larger than that of the $d=18$-dimensional
qudit code. Furthermore, comparing the $p$-expansions of (\ref{f1}) and (\ref%
{f3}) to the leading orders for $p\ll 1$, it follows that 
\begin{equation*}
\mathcal{F}^{\left( d=18\right) }\left( p\right) \overset{p\ll 1}{\approx }%
1-4p^{2}\leq 1-4p^{3}\overset{p\ll 1}{\approx }\mathcal{F}^{\left(
d=50\right) }\left( p\right) \text{.}
\end{equation*}%
From the above equation, it follows that the $d=50$-dimensional qudit code
outperforms the $d=18$-dimensional qudit code in the $p$-range where both
error correction schemes are effective as expected. Moreover, $\mathcal{F}%
^{\left( d=50\right) }\left( p\right) $ in (\ref{f3}) is a monotonic
decreasing function of the error probability parameter belonging in its
range of effectiveness.

\subsubsection{Asymmetric errors}

By taking probabilities as in Eqs.\eqref{piasym} and \eqref{pik} we obtain
for the $d=50$ qudit code 
\begin{align}
\mathcal{F}_{\text{asymmetric}}^{\left( d=50\right) }\left( p\right) &
=\left( 1-2p^{3}-2p^{4}-p^{5}\right) +\kappa^{3}\left( -2p^{3}+4p^{6}+
4p^{7}+2p^{8}\right)  \notag \\
& +\kappa^{4}\left( -2p^{4}+4p^{7}+4p^{8}+2 p^{9}\right) + \kappa ^{5}\left(
-p^{5}+2p^{8}+2p^{9}+p^{10}\right) \text{,}  \label{seconda}
\end{align}
Eqs.(\ref{f3}) and (\ref{seconda}) become, to the leading order in $p$ with $%
p\ll1$, 
\begin{equation*}
\mathcal{F}^{(d=50)}\left( p\right) \overset{p\ll1}{\approx}1-4p^{3},\quad%
\mathcal{F}_{\text{asymmetric}}^{(d=50)}\left( p\right) \overset{p\ll1}{
\approx} 1-2(1+\kappa^{3})p^{3}\text{.}
\end{equation*}
Also in this case, for $\kappa<1$ the presence of asymmetric Weyl errors
increases the performance of the correction scheme, while for $\kappa>1$ the
performance of the code is lowered.

Furthermore, comparing \eqref{prima} and \eqref{seconda} we get, to the
leading order in $p$ with $p\ll1$, 
\begin{equation*}
\mathcal{F}^{(d=18)}_{\text{asymmetric}}\left( p\right) \overset{p\ll 1}{%
\approx}1-2(1+\kappa^{2})p^{2},\quad\mathcal{F}_{\text{asymmetric}%
}^{(d=50)}\left( p\right) \overset{p\ll1}{ \approx} 1-2(1+\kappa^{3})p^{3}%
\text{.}
\end{equation*}
That is, the $d=50$ qudit code outperforms the $d=18$ one for any $\kappa \in%
[0,1/p]$.

%%%%%%%%%%%%%%%%%%%%%%%%%%%%%%%%%%%%%%%%%%%%%%%%%%%%%%%%%

\section{Comparison with block (stabilizer) codes}

In this Section, for the sake of comparison, we quantify the performance of
the standard five-qubit stabilizer code $[[5$, $1 $, $3]]$ \cite{laflamme,
bennett} and the seven-qubit stabilizer CSS (Calderbank-Shor-Steane) code $%
\left[ \left[ 7\text{, }1\text{, }3\right] \right] $ \cite{steane,
calderbank} on the tensor product of Weyl channels \eqref{Wch2} by means of
the entanglement fidelity \cite{schumy, mike}.

\subsection{The Five-Qubit Stabilizer Code}

\emph{Encoding}. The $\left[ \left[ 5\text{, }1\text{, }3\right] \right] $
code is the smallest single-error correcting quantum code. Of all quantum
codes that encode $1$ qubit and correct all single-qubit errors, the $\left[ %
\left[ 5\text{, }1\text{, }3\right] \right] $ is the most efficient,
saturating the quantum Hamming bound. It encodes $1$ qubit in $5$ qubits.
The cardinality of its stabilizer group $\mathcal{S}$ is $\left\vert 
\mathcal{S}\right\vert =2^{5-1}=16$ and the $5-1=4$ group generators are
given by \cite{gaitan}, 
\begin{equation*}
\left\{ X_{1}Z_{2}Z_{3}X_{4}\text{, }X_{2}Z_{3}Z_{4}X_{5}\text{ , }%
X_{1}X_{3}Z_{4}Z_{5}\text{, }Z_{1}X_{2}X_{4}Z_{5}\right\} \text{.}
\end{equation*}
The distance of the code is $d_{\mathcal{C}}=3$ and therefore the weight of
the smallest error $A_{l}^{\dagger}A_{k}$ that cannot be detected by the
code is $3$. Finally, we recall that it is a non-degenerate code since the
smallest weight for elements of $\mathcal{S}$ (other than identity) is $4 $
and therefore it is greater than the distance $d=3$. The encoding operation
for the $\left[ \left[ 5\text{, }1\text{, }3\right] \right] $ code is
characterized by, 
\begin{equation*}
\mathcal{H}_{2}\ni\left\vert 0\right\rangle \mapsto\left\vert
0_{L}\right\rangle \in\mathcal{H}_{2}^{\otimes5}\text{ and, }\mathcal{H}
_{2}\ni\left\vert 1\right\rangle \mapsto\left\vert 1_{L}\right\rangle \in%
\mathcal{H}_{2}^{\otimes5}\text{,}
\end{equation*}
where the codewords $\left\vert 0_{L}\right\rangle $ and $\left\vert
1_{L}\right\rangle $ are defined as \cite{gaitan}, 
\begin{equation*}
\left\vert 0\right\rangle \rightarrow\left\vert 0_{L}\right\rangle \overset{ 
\text{def}}{=}\frac{1}{4}\left[ 
\begin{array}{c}
\left\vert 00000\right\rangle +\left\vert 11000\right\rangle +\left\vert
01100\right\rangle +\left\vert 00110\right\rangle +\left\vert
00011\right\rangle +\left\vert 10001\right\rangle -\left\vert
01010\right\rangle -\left\vert 00101\right\rangle + \\ 
\\ 
-\left\vert 10010\right\rangle -\left\vert 01001\right\rangle -\left\vert
10100\right\rangle -\left\vert 11110\right\rangle -\left\vert
01111\right\rangle -\left\vert 10111\right\rangle -\left\vert
11011\right\rangle -\left\vert 11101\right\rangle%
\end{array}
\right] \text{,}  \label{5a}
\end{equation*}
and, 
\begin{equation*}
\left\vert 1\right\rangle \rightarrow\left\vert 1_{L}\right\rangle \overset{ 
\text{def}}{=}\frac{1}{4}\left[ 
\begin{array}{c}
\left\vert 11111\right\rangle +\left\vert 00111\right\rangle +\left\vert
10011\right\rangle +\left\vert 11001\right\rangle +\left\vert
11100\right\rangle +\left\vert 01110\right\rangle -\left\vert
10101\right\rangle -\left\vert 11010\right\rangle + \\ 
\\ 
-\left\vert 01101\right\rangle -\left\vert 10110\right\rangle -\left\vert
01011\right\rangle -\left\vert 00001\right\rangle -\left\vert
10000\right\rangle -\left\vert 01000\right\rangle -\left\vert
00100\right\rangle -\left\vert 00010\right\rangle%
\end{array}
\right] \text{,}  \label{5b}
\end{equation*}
respectively.

Then, the action of the channel $\Lambda^{(2)\otimes5}$ on $\rho \in%
\mathfrak{S}\left( \mathcal{H}_{2}^{\otimes5}\right) $ can be written as, 
\begin{equation}
\Lambda^{(2)\otimes5}(\rho)=\sum_{k=0}^{2^{10}-1}A_{k}\rho A_{k}^{\dagger }%
\text{,}  \label{nota}
\end{equation}
where, according to the notation of Section II, $A_{k}\propto
X(a_{1}a_{2}a_{3}a_{4}a_{5})Z(b_{1}b_{2}b_{3}b_{4}b_{5})$.

\emph{Correctability}. The sixteen weight zero and one quantum error
operators in (\ref{nota}) are given by,%
\begin{equation}
A_{0}=\sqrt{\tilde{p}_{0}}I_{1}\otimes I_{2}\otimes I_{3}\otimes
I_{4}\otimes I_{5}\text{, }A_{1}=\sqrt{\tilde{p}_{1}}X_{1}\otimes
I_{2}\otimes I_{3}\otimes I_{4}\otimes I_{5}\text{,..., }A_{15}=\sqrt{\tilde{%
p}_{15}}I_{1}\otimes I_{2}\otimes I_{3}\otimes I_{4}\otimes Z_{5}\text{,} 
\notag
\end{equation}%
where the coefficients $\tilde{p}_{l\text{ }}$ with $l=0$,..., $15$ can be
easily deduced from the above distribution of errors and Eq.\eqref{pparam}, $%
\tilde{p}_{0}=(1-p)^{10}$, $\tilde{p}_{1}=p(1-p)^{9}$, \ldots , $\tilde{p}%
_{6}=p^{2}(1-p)^{8}$, \ldots , $\tilde{p}_{15}=p(1-p)^{9}$.

It is straightforward, though tedious, to check that, for the above given
errors of weight zero and one we have: 
\begin{equation*}
S\left( A_{l}^{\dagger }A_{k}\right) \neq 0\text{, with }l\text{, }k\in
\left\{ 0\text{, }1\text{,..., }15\right\} \text{,}
\end{equation*}%
where $S\left( A_{l}^{\dagger }A_{k}\right) $ is the error syndrome of the
error operator $A_{l}^{\dagger }A_{k}$ defined according to \eqref{syndrom}
as, $S\left( A_{l}^{\dagger }A_{k}\right) \overset{\text{def}}{=}H^{\left[ %
\left[ 5,1,3\right] \right] }v_{A_{l}^{\dagger }A_{k}}$. The quantity $H^{%
\left[ \left[ 5,1,3\right] \right] }$ is the parity check matrix for the
five-qubit code while $v_{A_{l}^{\dagger }A_{k}}$ is the vector in the $10$%
-dimensional binary vector space $F_{2}^{10}$ corresponding to the error
operator $A_{l}^{\dagger }A_{k}$.

Hence, the set of correctable error operators is given by, 
\begin{equation*}
\mathcal{A}_{\text{correctable}}=\left\{ A_{0}\text{, } A_{1}\text{, }A_{2}%
\text{, }A_{3}\text{, } A_{4}\text{, }A_{5}\text{, }A_{6}\text{, } A_{7}%
\text{, }A_{8}\text{, }A_{9}\text{, } A_{10}\text{, }A_{11}\text{, }A_{12}%
\text{, } A_{13}\text{, }A_{14}\text{, }A_{15}\right\} \subseteq \mathcal{A}%
\text{,}
\end{equation*}
where the cardinality of $\mathcal{A}$ defining the channel in (\ref{nota})
equals $2^{10}$.

\emph{Recovery Operators}. All weight zero and one error operators satisfy
the error correction conditions \cite{knill02}, 
\begin{equation*}
P_{\mathcal{C}}A_{l}^{\dagger}A_{k}P_{\mathcal{C}}\propto P_{\mathcal{C}}%
\text{,}\quad l,k\in\left\{ 0\text{, }1\text{,..., }15\right\}
\label{detectability}
\end{equation*}
where $P_{\mathcal{C}}$ is the orthogonal projector operator ($P_{\mathcal{C}%
}=P_{\mathcal{C}}^{2}$ \ and, $P_{\mathcal{C}}=P_{\mathcal{C}}^{\dagger}$)
on the code space $\mathcal{C}^{\left( [[5,1,3]]\right) }$ defined as, 
\begin{equation*}
P_{\mathcal{C}}\overset{\text{def}}{=}\left\vert 0_{L}\right\rangle
\left\langle 0_{L}\right\vert +\left\vert 1_{L}\right\rangle \left\langle
1_{L}\right\vert \text{.}  \label{projector}
\end{equation*}
The two sixteen-dimensional orthogonal subspaces $\mathcal{V}^{0_{L}}$ and $%
\mathcal{V}^{1_{L}}$ of $\mathcal{H}_{2}^{\otimes5}$ generated by the action
of $\mathcal{A}_{\text{correctable}}^{\left[ \left[ 5,1,3\right] \right] }$
on $\left\vert 0_{L}\right\rangle $ and $\left\vert 1_{L}\right\rangle $ are
given by, 
\begin{equation*}
\mathcal{V}^{0_{L}}=Span\left\{ \left\vert v_{k}^{0_{L}}\right\rangle =\frac{%
A_{k}}{\sqrt{\tilde{p}_{k}}}\left\vert 0_{L}\right\rangle \right\} \text{,}%
\quad\text{and}\quad\mathcal{V}^{1_{L}}=Span\left\{ \left\vert
v_{k}^{1_{L}}\right\rangle =\frac{A_{k}}{\sqrt{\tilde{p}_{k}}}\left\vert
1_{L}\right\rangle \right\} \text{,}
\end{equation*}
with $k=0$, $1$,..., $15$. Notice that $\left\langle
v_{l}^{i_{L}}|v_{l^{\prime}}^{j_{L}}\right\rangle
=\delta_{ll^{\prime}}\delta_{ij}$ with $l$, $l^{\prime}\in\left\{ 0\text{, }1%
\text{,..., }15\right\} $ and $i$, $j\in\left\{ 0\text{, }1\right\} $.
Therefore, it follows that $\mathcal{V}^{0_{L}}\oplus\mathcal{V}^{1_{L}}=%
\mathcal{H}_{2}^{\otimes5}$. The recovery superoperator $\mathcal{R}%
\leftrightarrow\left\{ R_{l}\right\} $ with $l=1$ ,...,$16$ is defined by
means of \cite{KL}, 
\begin{equation*}
R_{l}=\left\vert 0_{L}\right\rangle \left\langle v_{l}^{0_{L}}\right\vert
+\left\vert 1_{L}\right\rangle \left\langle v_{l}^{1_{L}}\right\vert \text{.}
\end{equation*}
Finally, the composition of this recovery operation $\mathcal{R}$ with the
map $\Lambda^{\left( 2\right) \otimes5}\left( \rho\right) $ in (\ref{nota})
yields, 
\begin{equation}
\mathcal{R}\circ\Lambda^{(2)\otimes5}\left( \rho\right)
=\sum_{k=0}^{2^{10}-1}\sum\limits_{l=1}^{16}\left( R_{l}A_{k}\right)
\rho\left( R_{l}A_{k}\right) ^{\dagger}\text{.}  \label{r5}
\end{equation}

\begin{figure}[ptb]
\centering
\includegraphics[width=0.5\textwidth] {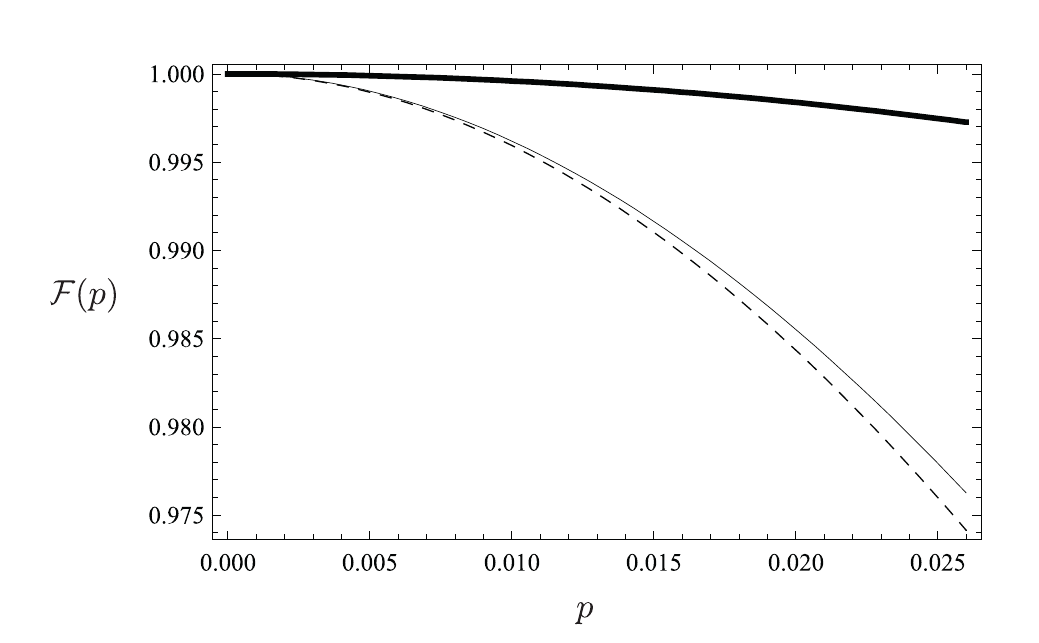}
\caption{ The quantities $\mathcal{F}^{(18)}$ (solid thick line), $\mathcal{F%
}^{[[5,1,3]]}$ (solid thin line), $\mathcal{F}^{[[7,1,3]]}$ (dash line), are
plotted vs. the error probability $p$ with $0\leq p\lesssim2.6\times10^{-2}$. 
Within the scale resolution of the graph the curve for $\mathcal{F}%
^{(50)}$ appears to coincide with the top horizontal axis.}
\label{fig1}
\end{figure}

\emph{Entanglement Fidelity}. Here we want to describe the action of $%
\mathcal{R\circ}\Lambda^{(2)\otimes5}$ in (\ref{r5}) restricted to the code
subspace $\mathcal{C}^{\left[ \left[ 5,1,3\right] \right] }$. Note that the
recovery operators can be expressed as, 
\begin{equation*}
R_{l+1}=R_{1}\frac{A_{l}}{\sqrt{\tilde{p}_{l}}}=\left( \left\vert
0_{L}\right\rangle \left\langle 0_{L}\right\vert +\left\vert
1_{L}\right\rangle \left\langle 1_{L}\right\vert \right) \frac{A_{l}}{\sqrt {%
\tilde{p}_{l}}}\text{,}
\end{equation*}
with $l\in\left\{ 0\text{,..., }15\right\} $. Recalling that in this case $%
A_{l}=A_{l}^{\dagger}$, it turns out that, 
\begin{equation*}
\left\langle i_{L}|R_{l+1}A_{k}|j_{L}\right\rangle =\frac{1}{\sqrt{\tilde {p}%
_{l}}}\left\langle i_{L}|0_{L}\right\rangle \left\langle
0_{L}|A_{l}^{\dagger}A_{k}|j_{L}\right\rangle +\frac{1}{\sqrt{\tilde{p}_{l}}}%
\left\langle i_{L}|1_{L}\right\rangle \left\langle
1_{L}|A_{l}^{\dagger}A_{k}|j_{L}\right\rangle \text{.}
\end{equation*}
We now need to compute the $2\times2$ matrix representation $\left[
R_{l}A_{k}\right] _{|\mathcal{C}}$ of each $R_{l}A_{k}$ with $l=0$,..., $15$
and $k=0$,..., $2^{10}-1$ where, 
\begin{equation*}
\left[ R_{l+1}A_{k}\right] _{|\mathcal{C}}\overset{\text{def}}{=}\left( 
\begin{array}{cc}
\left\langle 0_{L}|R_{l+1}A_{k}|0_{L}\right\rangle & \left\langle
0_{L}|R_{l+1}A_{k}|1_{L}\right\rangle \\ 
\left\langle 1_{L}|R_{l+1}A_{k}|0_{L}\right\rangle & \left\langle
1_{L}|R_{l+1}A_{k}|1_{L}\right\rangle%
\end{array}
\right) \text{.}
\end{equation*}
For $l$, $k=0$,.., $15$, we note that $\left[ R_{l+1}A_{k}\right] _{|%
\mathcal{C}}$ becomes, 
\begin{equation*}
\left[ R_{l+1}A_{k}\right] _{|\mathcal{C}}=\left( 
\begin{array}{cc}
\left\langle 0_{L}|A_{l}^{\dagger}A_{k}|0_{L}\right\rangle & 0 \\ 
0 & \left\langle 1_{L}|A_{l}^{\dagger}A_{k}|1_{L}\right\rangle%
\end{array}
\right) =\sqrt{\tilde{p}_{l}}\delta_{lk}\left( 
\begin{array}{cc}
1 & 0 \\ 
0 & 1%
\end{array}
\right) \text{,}
\end{equation*}
while for any pair $\left( l\text{, }k\right) $ with $l$, $=0$,..., $15$ and 
$k>15$, it follows that, 
\begin{equation*}
\left\langle 0_{L}|R_{l+1}A_{k}|0_{L}\right\rangle +\left\langle
1_{L}|R_{l+1}A_{k}|1_{L}\right\rangle =0\text{.}
\end{equation*}
We conclude that the only matrices $\left[ R_{l}A_{k}\right] _{|\mathcal{C}}$
with non-vanishing trace are given by, 
\begin{equation*}
\left[ R_{s}A_{s-1}\right] _{|\mathcal{C}}=\sqrt{\tilde{p}_{s-1}}\left( 
\begin{array}{cc}
1 & 0 \\ 
0 & 1%
\end{array}
\right) \text{, }
\end{equation*}
with $s=1$,.., $16$. Therefore, the entanglement fidelity \eqref{nfi} can be
written as 
\begin{equation*}
\mathcal{F}^{\left[ \left[ 5,1,3\right] \right] }\left( p\right) \overset{%
\text{def}}{=}\mathcal{F}^{\left[ \left[ 5,1,3\right] \right] }\left( \frac{1%
}{2}I_{2\times2}\text{, }\mathcal{R\circ}\Lambda^{(2)\otimes 5}\right) =%
\frac{1}{\left( 2\right) ^{2}}\sum_{k=0}^{2^{10}-1}\sum\limits_{l=1}^{16}%
\left\vert \text{tr}\left( \left[ R_{l}A_{k}\right] _{|\mathcal{C}}\right)
\right\vert ^{2}\text{,}  \label{fidel}
\end{equation*}
and results, 
\begin{equation}
\mathcal{F}^{\left[ \left[ 5,1,3\right] \right] }\left( p\right)
=1-40p^{2}+200p^{3}-490p^{4}+728p^{5}-700p^{6}+440p^{7}+175p^{8}+40p^{9}-4p^{10}%
\text{.}  \label{f4}
\end{equation}
We remark that this error correction scheme is effective as long as the
failure probability $1-\mathcal{F}^{\left[ \left[ 5,1,3\right] \right]
}\left( p\right) $ is strictly smaller than the error probability $p$. This
implies that the five-qubit code is effective only when $0\leq p\lesssim
2.9\times10^{-2}$. Finally, we emphasize that this block-encoding scheme is
less efficient than the previously-mentioned qudit-encoding schemes as it
appears in Figure \ref{fig1}.

\subsubsection{Asymmetric errors}

By taking probabilities as in Eqs.\eqref{piasym} and \eqref{pik} we obtain
for the five-qubit code 
\begin{align}
\mathcal{F}_{\text{asymmetric}}^{\left[ \left[ 5,1,3\right] \right] }\left(
p\right) & =1-10p^{2}+20p^{3}-15p^{4}+4p^{5}  \notag \\
& +\kappa\left( -20p^{2}+80p^{3}-\allowbreak120p^{4}+80p^{5}-20p^{6}\right) 
\notag \\
& +\kappa^{2}\left(
-10p^{2}+80p^{3}-220p^{4}+280p^{5}-170p^{6}+40p^{7}\right)  \notag \\
& +\kappa^{3}\left( \allowbreak20p^{3}-120p^{4}+280p^{5}-320\allowbreak
p^{6}+180p^{7}\allowbreak-40p^{8}\right)  \notag \\
& +\kappa^{4}\left(
-\allowbreak15p^{4}+80p^{5}-170p^{6}+180p^{7}-\allowbreak95p^{8}+20p^{9}%
\right)  \notag \\
& +\kappa^{5}\left(
4p^{5}-20p^{6}+40p^{7}-40p^{8}+20p^{9}-4p^{10}\allowbreak\right) \text{.}
\label{Fasym5}
\end{align}
We stress that unlike the finding uncovered in \cite{carlo-PRA}, asymmetries
in the considered Weyl noisy channel do affect the performance of the
five-qubit code quantified in terms of the entanglement fidelity. This
difference is ultimately a consequence of the fact that while in \cite%
{carlo-PRA} it is assumed error probabilities $p_{X}$, $p_{Y}$ and $p_{Z}$
all \emph{linear} in the error probability $p$ (although weighted with
different coefficients $\alpha_{X}$, $\alpha_{Y}$ and $\alpha_{Z}$ with $%
\alpha_{X}+\alpha_{Y}+\alpha_{Z}=1$), here we assumed $p_{Y}$ \emph{quadratic%
} in $p$ while keeping both $p_{X}$ and $p_{Z}$ \emph{linear} in $p$.

From Eqs.\eqref{f4} and \eqref{Fasym5} we obtain, to the leading order in $p$
with $p\ll1$, 
\begin{equation*}
\mathcal{F}^{\left[ \left[ 5,1,3\right] \right] }\left( p\right) \overset{%
p\ll1}{\approx} 1-40p^{2},\quad\mathcal{F}_{\text{asymmetric}%
}^{[[5,1,3]]}\left( p\right) \overset{p\ll1}{\approx} 1-10(1+\kappa
)^{2}p^{2} \text{.}
\end{equation*}
Hence, also in this case, the presence of asymmetric errors increases the
performance of the correction scheme for $\kappa<1$, while for $\kappa>1$
the performance of the code is lowered.

\subsection{The Seven-Qubit Stabilizer Code}

The $\left[ \left[ 7\text{, }1\text{, }3\right] \right] $-CSS code encodes $%
1 $ qubit in $7$ qubits. The cardinality of its stabilizer group $\mathcal{S}
$ is $\left\vert \mathcal{S}\right\vert =2^{7-1}=64$ and the set of $7-1=6$
group generators is given by \cite{gaitan}, 
\begin{equation*}
\left\{ X_{4}X_{5}X_{6}X_{7}\text{, }X_{2}X_{3}X_{6}X_{7}\text{ , }%
X_{1}X_{3}X_{5}X_{7}\text{, }Z_{4}Z_{5}Z_{6}Z_{7}\text{, }%
Z_{2}Z_{3}Z_{6}Z_{7}\text{, }Z_{1}Z_{3}Z_{5}Z_{7}\right\} \text{.}
\end{equation*}%
The distance of the code is $d_{\mathcal{C}}=3$ and therefore the weight of
the smallest error $A_{l}^{\dagger }A_{k}$ that cannot be detected by the
code is $3$. Finally, we recall that it is a non-degenerate code since the
smallest weight for elements of $\mathcal{S}$ (other than identity) is $4$
and therefore it is greater than the distance $d_{\mathcal{C}}=3$. The
encoding for the $\left[ \left[ 7\text{, }1\text{, }3\right] \right] $ code
is given by \cite{gaitan}, 
\begin{equation*}
\left\vert 0\right\rangle \rightarrow \left\vert 0_{L}\right\rangle =\frac{1%
}{\left( \sqrt{2}\right) ^{3}}\left[ 
\begin{array}{c}
\left\vert 0000000\right\rangle +\left\vert 0110011\right\rangle +\left\vert
1010101\right\rangle +\left\vert 1100110\right\rangle + \\ 
\\ 
+\left\vert 0001111\right\rangle +\left\vert 0111100\right\rangle
+\left\vert 1011010\right\rangle +\left\vert 1101001\right\rangle%
\end{array}%
\right] \text{,}
\end{equation*}%
and, 
\begin{equation*}
\left\vert 1\right\rangle \rightarrow \left\vert 1_{L}\right\rangle =\frac{1%
}{\left( \sqrt{2}\right) ^{3}}\left[ 
\begin{array}{c}
\left\vert 1111111\right\rangle +\left\vert 1001100\right\rangle +\left\vert
0101010\right\rangle +\left\vert 0011001\right\rangle + \\ 
\\ 
+\left\vert 1110000\right\rangle +\left\vert 1000011\right\rangle
+\left\vert 0100101\right\rangle +\left\vert 0010110\right\rangle%
\end{array}%
\right] \text{.}
\end{equation*}%
Following the same line of reasoning of the previous Subsection we can
compute the correctable errors by means of $H^{\left[ \left[ 7,1,3\right] %
\right] }$, the parity check matrix for the seven-qubit code. Finally, after
determining the recovery operators, it can be shown that the entanglement
fidelity reads, 
\begin{align}
\mathcal{F}^{\left[ \left[ 7,1,3\right] \right] }\left( p\right) &
=1\bigskip -42p^{2}+140p^{3}+231p^{4}-2772p^{5}+9240p^{6}-\allowbreak
18\,216p^{7}+24\,255p^{8}-22\,792p^{9}  \notag \\
& +15\,246p^{10}-\allowbreak 7140p^{11}+2233p^{12}-420p^{13}+36p^{14}\text{.}
\label{f5}
\end{align}%
Observe that the seven-qubit code is effective for $0\leq p\lesssim
2.6\times 10^{-2}$. Comparing the $p$-expansions of (\ref{f4}) and (\ref{f5}%
) to the leading orders for $p\ll 1$, it follows that 
\begin{equation*}
\mathcal{F}^{\left[ \left[ 5,1,3\right] \right] }\left( p\right) \overset{%
p\ll 1}{\approx }1-40p^{2}\geq 1-42p^{2}\overset{p\ll 1}{\approx }\mathcal{F}%
^{\left[ \left[ 7,1,3\right] \right] }\left( p\right) \text{,}
\end{equation*}%
and, in addition, the $p$-range of applicability of the five-qubit code is
larger than that of the seven-qubit code. Thus, for the symmetric Weyl
channel considered, the five-qubit code outperforms the seven-qubit code.
However, we shall see that this ordering does not hold when considering
asymmetric scenarios.

\subsubsection{Asymmetric errors}

In the asymmetric scenario of Eqs.\eqref{piasym} and \eqref{pik} we obtain
for the seven-qubit code 
\begin{align}
\mathcal{F}_{\text{asymmetric}}^{\left[ \left[ 7,1,3\right] \right] }\left(
p\right) & =1-21p^{2}+70p^{3}-105p^{4}+84p^{5}-35p^{6}+6p^{7}  \notag \\
& +\kappa\left(
-\allowbreak21p^{2}+126p^{3}-315p^{4}+420p^{5}-315p^{6}%
\allowbreak+126p^{7}-21p^{8}\right)  \notag \\
& +\kappa^{2}\left( -21p^{3}+126\allowbreak
p^{4}-315p^{5}+420p^{6}-\allowbreak315p^{7}+126p^{8}-21p^{9}\right)  \notag
\\
& +\kappa^{3}\left( -35p^{3}+420p^{4}-1785\allowbreak
p^{5}+3850p^{6}-4725p^{7}\allowbreak+3360\allowbreak
p^{8}-1295p^{9}+210p^{10}\right)  \notag \\
& +\kappa^{4}\left( 105p^{4}-1050p^{5}+4095p^{6}-8400p^{7}+9975\allowbreak
p^{8}-\allowbreak6930p^{9}+2625p^{10}-420p^{11}\right)  \notag \\
& +\kappa^{5}\left(
-126p^{5}+1155p^{6}-4284p^{7}+8505p^{8}-9870p^{9}+6741p^{10}-2520p^{11}+399p^{12}\right)
\notag \\
& +\kappa^{6}\left( 70p^{6}-609p^{7}+2184p^{8}\allowbreak-4235\allowbreak
p^{9}+\allowbreak4830p^{10}-3255\allowbreak p^{11}+1204p^{12}\allowbreak
-189p^{13}\right)  \notag \\
& +\kappa^{7}\left(
-15p^{7}+126p^{8}-441p^{9}+840p^{10}-945p^{11}+630p^{12}-231p^{13}+36%
\allowbreak p^{14}\right) \text{.}  \label{Fasym7}
\end{align}
From Eqs.(\ref{f5}) and (\ref{Fasym7}) it follows, to the leading order in $%
p $ with $p\ll1$, 
\begin{equation*}
\mathcal{F}^{[[7,1,3]]}\left( p\right) \overset{p\ll1}{\approx}%
1-42p^{2},\quad\mathcal{F}_{\text{asymmetric}}^{\left[ \left[ 7,1,3\right] %
\right] }\left( p\right) \overset{p\ll1}{ \approx} 1-21(1+\kappa)p^{2} \text{%
.}
\end{equation*}
Once again, for $\kappa<1$ it results that the presence of asymmetric errors
increases the performance of the correction scheme, while for $\kappa>1$ the
performance of the code is lowered.

Furthermore, by comparing \eqref{Fasym5} and \eqref{Fasym7} it follows, to
the leading order in $p$ with $p\ll1$, 
\begin{align}
& \mathcal{F}_{\text{asymmetric}}^{\left[ \left[ 5,1,3\right] \right]
}\left( p\right) \overset{p\ll1}{\approx}1-10(1+\kappa)^{2}p^{2} >
1-21(1+\kappa)p^{2}\overset{p\ll1}{ \approx}\mathcal{F}_{\text{asymmetric}}^{%
\left[ \left[ 7,1,3\right] \right] }\left( p\right) \text{,}\quad\kappa<1.1 
\notag \\
& \mathcal{F}_{\text{asymmetric}}^{\left[ \left[ 5,1,3\right] \right]
}\left( p\right) \overset{p\ll1}{\approx}1-10(1+\kappa)^{2}p^{2} <
1-21(1+\kappa)p^{2}\overset{p\ll1}{ \approx}\mathcal{F}_{\text{asymmetric}}^{%
\left[ \left[ 7,1,3\right] \right] }\left( p\right) \text{,}\quad\kappa>1.1 
\notag  \label{7vs5}
\end{align}
Thus, with respect to the noise model discussed in Ref.\cite{carlo-PRA}, we
conclude that here the comparison between five-qubit code and seven-qubit
code is slightly more involved.

\section{Final Remarks}

In this article, we discussed how to protect a qubit embedded into a qudit
from both amplitude and phase errors occurring in the discrete phase space.
A code has been devised using stabilizer formalism and its performances
compared with those of common block codes for a general Weyl noisy quantum
channel allowing symmetric and asymmetric error probabilities.

Specifically we have considered the $d=18$ and $d=50$ qudit stabilizer codes
together with five and the CSS (Calderbank-Steane-Shor) seven-qubit quantum
stabilizer codes. The performances of these codes were quantified by means
of the entanglement fidelity as function of the error probability.

We uncovered that qudit codes have an enormously wider (by approximately an
order of magnitude) range of applicability in the error probability.
Furthermore, already the $d=18$ qudit code outperforms the five and
seven-qubit block codes for symmetric errors (see Fig.\ref{fig1}). Our
theoretical analysis leads to the conclusion that the qudit codes with $d=18$%
\ and $d=50$\ outperform the common five and CSS seven-qubit stabilizer
codes. This in principle allows one to save space resources (since $%
d=18<\dim _{\mathbb{C}}H_{2}^{\otimes 5}=32<<\dim _{\mathbb{C}%
}H_{2}^{\otimes 7}=128$), however one should also account for the
difficulties in implementing qudit systems, an issue that seems to be
nontrivial and not quite settled yet. For an overview of the experiments
performed for producing quantum optical qudits, we refer to \cite{genovese}. 
\begin{figure}[t]
\begin{center}
\includegraphics[width=0.5\textwidth]{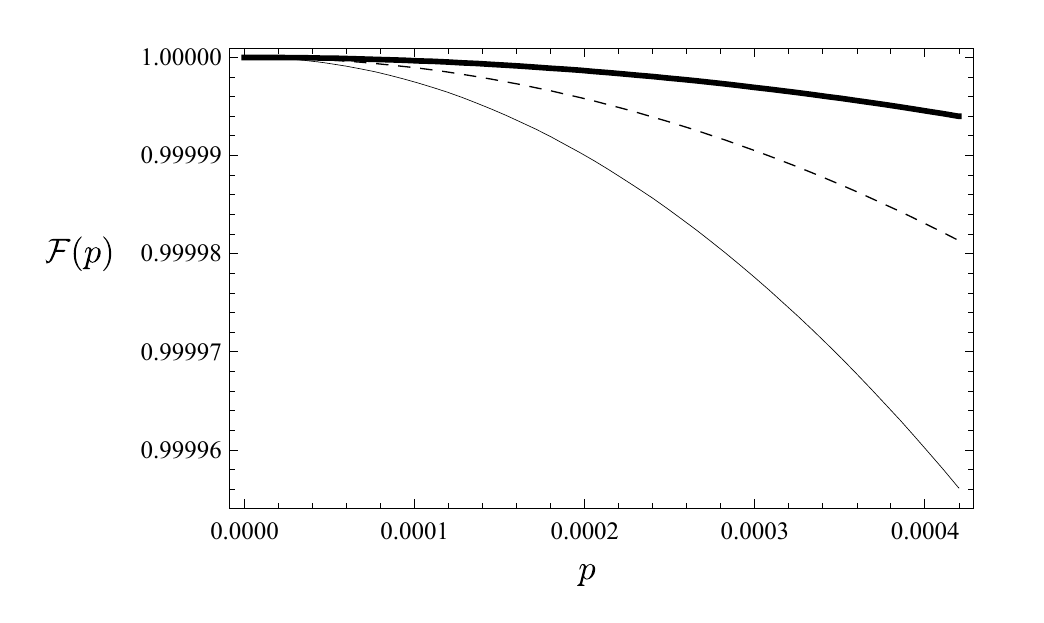}\includegraphics[width=0.53%
\textwidth] {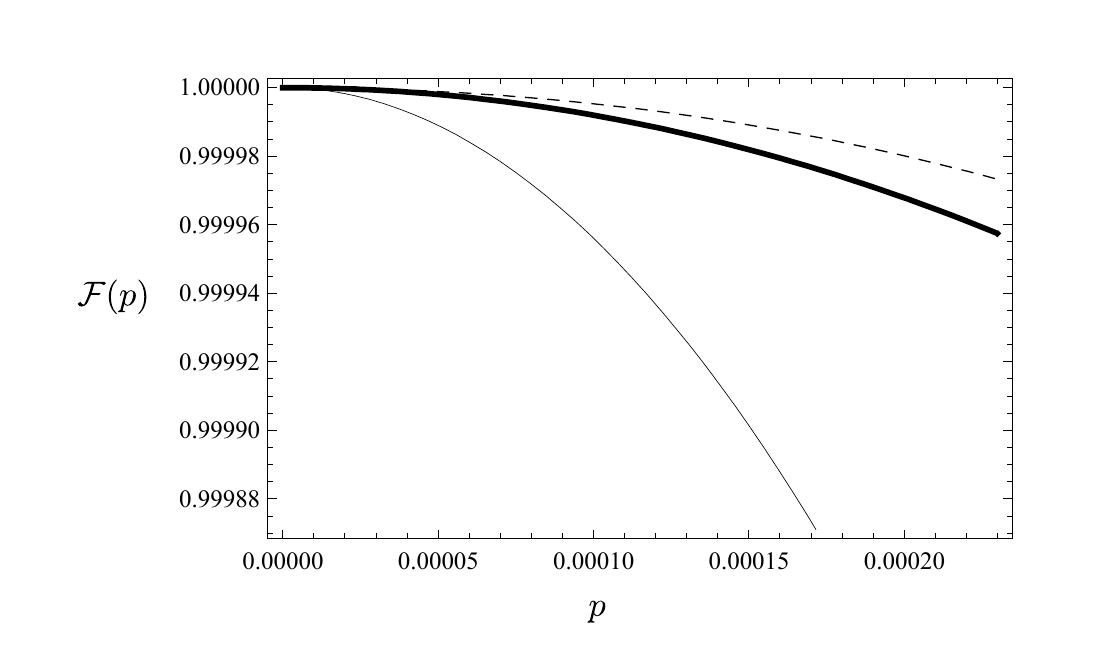}
\end{center}
\caption{ The quantities $\mathcal{F}_{\text{asymmetric}}^{(18)}$ (solid
thick line), $\mathcal{F}_{\text{asymmetric}}^{[[5,1,3]]}$ (solid thin
line), $\mathcal{F}_{\text{asymmetric}}^{[[7,1,3]]}$ (dash line), are
plotted vs. the error probability $p$ for $\protect\kappa =4<11.34$ (left)
and for $\protect\kappa =20>11.34$ (right). Within the scale
resolution of the graphs the curves for $\mathcal{F}_{\text{asymmetric}%
}^{(50)}$ appear to coincide with the top horizontal axes.}
\label{fig2lr}
\end{figure}

The performance of qudit code are also robust against asymmetries in errors'
probabilities. In fact, restricting our analysis to $\kappa>1.1$, it results
that the $d=18$ qudit code outperforms the seven-qubit code until strong
asymmetries come into play, as can be seen by comparing \eqref{prima} with %
\eqref{Fasym7} to the leading order in $p$ with $p\ll1$, 
\begin{eqnarray}
& \mathcal{F}_{\text{asymmetric}}^{\left[ \left[ 7,1,3\right] \right]
}\left( p\right) \overset{p\ll1}{\approx}1-21(1+\kappa)p^{2} <
1-2(1+\kappa^{2})p^{2}\overset{p\ll1}{ \approx}\mathcal{F}_{\text{asymmetric}%
}^{(d=18) }\left( p\right) \text{,}\quad\kappa<\frac{21+\sqrt{593}}{4}%
\approx11.34  \notag \\
& \mathcal{F}_{\text{asymmetric}}^{\left[ \left[ 7,1,3\right] \right]
}\left( p\right) \overset{p\ll1}{\approx}1-21(1+\kappa)p^{2} >
1-2(1+\kappa^{2})p^{2}\overset{p\ll1}{ \approx}\mathcal{F}_{\text{asymmetric}%
}^{(d=18) }\left( p\right) \text{,}\quad\kappa>\frac{21+\sqrt{593}}{4}%
\approx11.34  \notag  \label{7vs18}
\end{eqnarray}

Comparative results for the various codes performances in presence of
asymmetries are graphically represented in Figure \ref{fig2lr}.

The different uncovered behaviors in the four error correcting schemes
employed in this article can be ascribed to the fact that the errors in $%
\mathcal{P}_{\mathcal{H}_{2}^{\otimes5}}$ (or $\mathcal{P}_{\mathcal{H}%
_{2}^{\otimes7}}$) are fundamentally different from those in $\mathcal{P}_{%
\mathcal{H}_{18}}$ and $\mathcal{P}_{\mathcal{H}_{50}}$.

Finally, it could be interesting to consider the presence of correlations
between $X$ and $Z$ errors in the qudit code. These can be introduced as
follow 
\begin{equation*}
\pi\left( n\text{, }m\right) \overset{ \text{def}}{=}\left( 1-\mu\right)
\pi\left( n\right) \pi\left( m\right) +\mu\delta_{n\text{, }m}\pi\left(
m\right) \text{,}
\end{equation*}
where $\mu$ with $0\leq\mu\leq1$ represent the degree of correlation.
Following the very same line of reasoning provided in Section IV, it can be
shown that for instance the entanglement fidelity becomes, 
\begin{equation*}
\mathcal{F}_{\text{corr}}^{\left( d=18\right) }\left( p\right) =\left(
1-4p^{2}-2p^{3}+4p^{4}+4p^{5}+p^{6}\right) +\mu\left(
2p^{2}+p^{3}-4p^{4}\allowbreak-4p^{5}-p^{6}\right) \text{.}  \label{c18}
\end{equation*}
It then results $\left( \partial\mathcal{F}^{\left( d=18\right) }_{\text{corr%
}}/\partial\mu\right) _{p=const.} \overset{p\ll1}{\approx}2p^{2} \geq0$,
that is memory effects lead to better performances. The reason is that in
the limit of very strong correlations $\mu\to1$, only one type of error
(namely $Y=XZ$) takes place. As such, this case shows similarities with the
case of asymmetric errors with $\kappa\to0$.

In conclusion, we are strongly motivated by our investigation to believe
that encoding a qubit into a qudit can be a useful approach in quantum
coding. We are aware of the difficulties in realizing and controlling qudit
systems even of low dimensionality, however we have witnessed a lot of
progress along this direction recently. Quantum optical qudits can be
generated by means of experimental schemes based upon interferometric
set-ups, orbital angular momentum entanglement and, biphoton polarization 
\cite{genovese}. For instance, in the interferometric scheme employed in 
\cite{te}, high symmetry and maximally entangled qutrits are realized with a
fidelity up to $0.985$\ as the superposition state of the three possible
paths of a single photon in a three-arms interferometer. Therefore, the
realization of the discussed qudit codes seems not futureless.

\begin{acknowledgments}
This work has been supported by the European Commission's Seventh Framework
Programme (FP7/2007--2013) under grant agreement no. 213681. C. C. thanks
Hussain Zaidi for technical assistance and Peter van Loock for discussions
about quantum codes.
\end{acknowledgments}

%%%%%%%%%%%%%%%%%%%%%%%%%%%%%%%%%%%%%%%%%%%%%%

\end{document}